\input epsf.tex

\documentclass[preprint,pra,aps,showpacs,preprintnumbers,amsmath,amssymb,eqsecnum]{revtex4}

\usepackage{graphicx}

\def\a{{\alpha}}

\def\e{{\epsilon}}
\def\eps{{\epsilon}}

\def\t{\tau}

\def\a{\alpha}

\def\half{\frac{1}{2}}

\newcommand\beq{\begin{equation}}
\newcommand\eeq{\end{equation}}
\newcommand\bea{\begin{eqnarray}}
\newcommand\eea {\end{eqnarray}}

\begin{document}


\title{On the Relationship Between Complex Potentials and Strings of Projection Operators}

\author{J.J.Halliwell}

\author{J.M.Yearsley}

\affiliation{Blackett Laboratory \\ Imperial College \\ London SW7
2BZ \\ UK }

\date{\today}

\begin{abstract}
It is of interest in a variety of contexts, and in particular in the arrival time problem,
to consider the quantum state obtained through unitary evolution
of an initial state regularly interspersed with periodic projections onto the positive $x$-axis (pulsed
measurements).
Echanobe, del Campo and Muga have given a compelling but heuristic argument
that the state thus obtained is approximately
equivalent to the state obtained by evolving in the presence of a certain complex
potential of step-function form. In this paper, with the help of the path decomposition
expansion of the associated propagators, we give a detailed derivation of this approximate equivalence.
The propagator for the complex potential is known so the bulk of the derivation consists
of an approximate evaluation of the propagator for the free particle interspersed
with periodic position projections.  This approximate equivalence may be used to show that
to produce significant reflection, the projections must act at time spacing less than
$ \hbar / E$, where $E$ is the energy scale of the initial state.

\end{abstract}

\maketitle

\section{Introduction}

A number of physically interesting situations in quantum theory
concern the question of what happens to an initial quantum state $
| \psi \rangle $ which is acted on by a sequence of projection
operators interspersed with unitary evolution for time $\epsilon$,
\beq
| \psi_P (\tau) \rangle = e^{-iH \epsilon}
Pe^{-i H \epsilon}...P e^{-i H \epsilon} | \psi \rangle
\label{1.1}
\eeq
(Here, there are $n$ projection operators and $ \tau = (n+1) \epsilon$
and we use units in which $\hbar =1 $).
Such an object describes pulsed measurements but it also crops up in
the decoherent histories approach to quantum theory, where the right-hand side
is the amplitude for a quantum-mechanical history
\cite{GeH1,GeH2,Gri,Omn,Hal2,DoH}.

We are interested in the specific case of a free particle with $P$ taken to be
the projection onto the positive $x$-axis, $P = \theta ( \hat x)$. For
sufficiently small $\epsilon$, Eq.(\ref{1.1}) is then a candidate for
the amplitude to remain in $x>0$ during the time interval $[0, \tau]$,
an object that is of interest in the arrival time problem \cite{time,All,HaYe1,HaYe2}.

It is of interest
to explore the properties of this amplitude for a range of values
of the time spacing $\epsilon$. It is known that as $\epsilon \rightarrow 0$,
we approach the Zeno limit, in which the state becomes entirely
confined to the Hilbert subspace
of states with support only in $x>0$, so that an incoming wave packet
from the right is totally reflected \cite{Zeno,Wall,Sch2}.
However, it is of greater physical
interest to explore the regime of non-zero $\epsilon$, in which the system
is monitored sufficiently well to get some idea of whether the particle is in $x>0$,
yet not monitored so much that an incoming state is significantly
reflected at $x=0$. An important question in this regime is to determine the
value of $ \eps $ for which reflection becomes significant. For an initial
state with energy width $\Delta H$, a timescale held to be significant
is the Zeno time,
\beq
t_Z = \frac {1} { \Delta H}
\eeq
which is the timescale on which the state becomes significantly different
from its initial value under unitary evolution \cite{Zeno}.
For a wave
packet of momentum $p$ and spatial width $\sigma$, the Zeno time is of order $m \sigma / p$
which is the timescale on which the wave packet crosses the origin.
This indicates that the Zeno time for wave packets
is an essentially classical timescale and, in Eq.(\ref{1.1}), relates only
to the rate of removal of probability through projection. By contrast,
reflection in Eq.(\ref{1.1}) arises as a result of the increase in uncertainty in momentum
resulting from position projection, an obviously quantum process, so one would expect it to have a different
timescale, which could be much shorter than then Zeno time. It would be of
interest to compute this timescale. One reason it is important is that
there appears to be interesting physics very close to the Zeno limit \cite{Ech,Muga2,Hal4}.

A significant result in this area is due to Echanobe, del Campo and Muga, who claimed
that for finite $\epsilon$ the string of projection operators in Eq.(\ref{1.1})
is approximately equivalent to evolution in the presence of a complex potential
\cite{Ech}.
That is,
\beq
e^{ - i H \eps} Pe^{-i H \epsilon} \cdots P e^{-i H \eps}
\approx \exp \left( - i H \tau - V_0 \theta ( -x ) \tau \right)
\label{1.2}
\eeq
They assert that this result is valid if, for a given $\eps$, $V_0$ is chosen such that
two conditions
\bea
V_0 \epsilon  & \gg & 1
\label{1.3} \\
V_0 & \gg & \Delta H
\label{1.4}
\eea
are satisfied.
This is a very useful result since Eq.(\ref{1.1}) is not easy to evaluate analytically
but the Schr\"odinger equation with a complex step potential in Eq.(\ref{1.2}) can be
solved straightforwardly. Furthermore, such complex potentials have been studied extensively
in the literature and can often be linked to particular detection methods
\cite{complex}.

Given the connection Eq.(\ref{1.2}), one can determine the conditions under which
reflection becomes important. Known results on scattering with the complex
potential in Eq.(\ref{1.2}) show that, for an incoming state with energy scale $E$,
reflection becomes significant when $V_0 > E$ \cite{All, HaYe1}. Reflection is avoided,
therefore, when $ V_0 \ll E $, or equivalently, from Eq.(\ref{1.3}), when
\beq
\eps \gg \frac {1} {E}
\label{1.6}
\eeq
This is much less than the Zeno time for a state strongly peaked in energy.
There is therefore an interesting regime, namely
\beq
\frac{1} {E} \ \ll \ \eps \ \ll \ \frac {1} { \Delta H}
\label{1.7}
\eeq
in which the projections in Eq.(\ref{1.1}) are sufficiently frequent
to have a significant effect on the system, yet not so frequent that there is significant reflection.

Eq.(\ref{1.6}) is a very useful result, but it has been derived on the basis
of the claimed approximate
relationship Eq.(\ref{1.2}). The derivation of Eq.(\ref{1.2})
given by Echanobe et al
is very plausible (and was also hinted at by Allcock \cite{All}), but it
is rather heuristic, and the deduced connection
Eq.(\ref{1.3}) between $V_0$ and $\eps$ is rather loose.
There is therefore considerable scope for
a more detailed and substantial derivation.

The purpose of this paper is to give a more substantial derivation of the equivalence
Eq.(\ref{1.2}) and to deduce a more precise relationship between $ V_0$ and $\eps$.
We will do this by computing the configuration space propagators associated
with each side of Eq.(\ref{1.2}) and show that they are approximately equal in
certain regimes. The propagator associated with the complex potential is in fact
known already, so the bulk of the work consists of an approximate evaluation of
the propagator associated with pulsed measurements, the left-hand side of Eq.(\ref{1.2}).

This is certainly not a rigorous mathematical proof of Eq.(\ref{1.2}), involving operator norms,
error bounds and the like, but a theoretical physicists style of proof involving
the approximate evaluation of propagators. A rigorous proof would certainly be of interest to construct
and the work described in this paper may give some hints in that direction.

We begin in Section 2 with a brief summary of the derivation of Echanobe et al,
with a small extension of it that turns out to be important and yields an equality
relating $V_0$ and $\eps$, thereby improving on Eq.(\ref{1.3}).
We then in Section 3 give a detailed formulation of the problem and how we solve it.
The key idea is to use the path decomposition expansion in which the propagators associated with
each side of Eq.(\ref{1.2}) are factored across the surface $x=0$
\cite{PDX,HaOr,Hal3}. The problem
of proving the equivalence Eq.(\ref{1.2}) thereby reduces to proving it for propagation
between points at $x=0$ for different times. Since the propagator for the complex potential
is known, the main work is to evaluate this propagator for pulsed measurements. This is actually
rather difficult to do directly, but a good approximate analytic expression can be obtained for it
by attacking the problem from a number of different angles. This is described in
Sections 4, 5, 6 and 7. In Section 8
we give a detailed discussion of the timescales involved for the approximations
to be valid.
We summarize and conclude in Section 9.

\section{Review and Extension of Earlier Work}

We first review and extend the derivation of Echanobe et al \cite{Ech}. They first note that
\beq
 \exp \left( -  V_0  \theta ( - \hat x ) \epsilon  \right)
 = P + e^{ - V_0 \eps } \bar P
\label{2.AA}
 \eeq
where, recall $P = \theta ( \hat x ) $ and  $\bar P = 1 - P $. It follows that
\beq
P = \theta (\hat x ) \approx  \exp \left( -  V_0  \theta ( - \hat x ) \epsilon  \right)
\label{2.A}
\eeq
as long as the parameter
\beq
\alpha = V_ 0 \epsilon
\label{2.B}
\eeq
is sufficiently large that
\beq
e^{ - \alpha } \ll 1
\label{2.BB}
\eeq
Eq.(\ref{1.2}) then follows from the approximate equivalence,
\beq
\exp \left( - i H \epsilon \right)
\ \exp \left( -  V_0  \theta ( - \hat x ) \epsilon  \right)
\approx
\exp \left( - i H \epsilon  -  V_0  \theta ( - \hat x ) \epsilon  \right)
\eeq
which will hold as long as
\beq
V_0 \epsilon^2 |\langle [H, \theta (-\hat x) ] \rangle| \ll 1
\label{2.C}
\eeq
where the average is taken in the initial state.
Echanobe et al put an upper bound on the left-hand side using the Schr\"odinger-Robertson
inequality and Eq.(\ref{2.C}) may then be written in either of the two equivalent forms
\beq
\alpha^2 \Delta H \ll V_0
\label{2.D}
\eeq
or
\beq
\alpha \eps \ll \frac {1} { \Delta H}
\label{2.DD}
\eeq
which implies that the time between projections is much less than the Zeno time,
the typical timescale on which the state undergoes significant change \cite{Zeno}.
(The conditions Eqs.(\ref{2.B}), (\ref{2.BB}), (\ref{2.D}) are a more precise version of the originally stated conditions
Eqs.(\ref{1.3}), (\ref{1.4})).

Since the parameter $\alpha$ need only satisfy an {\it inequality}, Eq.(\ref{2.BB}), the
relationship Eq.(\ref{2.B}) between $\eps$ and $V_0$ is not uniquely determined. However, it turns
out that the above derivation can be extended somewhat to give an {\it equality}
between $\eps$ and $V_0$. This turns out to be relevant to the more substantial derivation
given in the rest of this paper.

The above result may be written,
\beq
\exp \left( - i H \epsilon  -  V_0  \theta ( - \hat x ) \epsilon  \right)
\approx \exp \left( - i H \epsilon \right) P
\label{2.E}
\eeq
under the conditions given above. Now suppose $\alpha $ is an integer and
write $ \eps = \alpha \eps'$, where
\beq
V_0 \eps' =1
\label{2.EE}
\eeq
Then, since
$P= P^2 $, we may approximate Eq.(\ref{2.E}) by
\bea
\exp \left( - i H \epsilon  -  V_0  \theta ( - \hat x ) \epsilon  \right)
& \approx &
e^{ - i H \eps'} Pe^{-i H \epsilon'} P \cdots e^{-i H \eps'} P
\nonumber \\
&=& \left( e^{ - i H \eps'} P \right)^{\alpha}
\label{2.F}
\eea
as long as the contribution from the commutator terms between $ e^{-iH \eps'}$ and $P$
is sufficiently
small. There will be of order $\alpha^2$ such terms, hence the error in this
approximation is of order $ \alpha^2 \eps' \Delta H $ and from Eq.(\ref{2.D}),
this error is much less than $1$. The key point here is that even the longer timescale
$\eps$ is still much less than the Zeno time and, since nothing changes on this
timescale, there is essentially no difference between Eqs.(\ref{2.E}) and (\ref{2.F}).

We therefore see that the desired result Eq.(\ref{1.2}) actually holds for much
smaller time steps, defined by the equality Eq.(\ref{2.EE}). This is
different to the original claim of Echanobe et al, since they require the
inequality Eq.(\ref{1.3}). However, the above argument shows that the restriction
Eq.(\ref{1.3}) is in fact stronger than necessary and in the following pages
our more detailed derivation will show that
Eq.(\ref{1.2}) does indeed hold with a timespacing of order $ 1 / V_0 $.

\section{Detailed Formulation of the Problem}

In this paper, we will prove the relationship Eq.(\ref{1.2}) in a much more substantial
way by proving the approximate equivalence of the propagators
\bea
g_V ( x_1, \tau | x_0, 0 ) &=& \langle x_1 |   \exp \left( - i H \tau - V_0 \theta ( -x ) \tau \right)
| x_0 \rangle
\label{2.2}
\\
g_P ( x_1, \tau | x_0, 0 ) &=& \langle x_1 | e^{-iH \epsilon_{n}} Pe^{-i H \epsilon}
...e^{-i H \epsilon}P e^{-i H \eps_0} | x_0 \rangle
\label{2.1}
\eea
for some relationship between the parameters $\eps$ and $V_0$, to be determined.
Note that in Eq.(\ref{2.1}), we have chosen the initial and final time spacings
to be $\eps_0$ and $\eps_n $, with
\beq
\tau = (n-1) \eps + \eps_0 + \eps_n
\label{2.3}
\eeq
This turns
out to be necessary for the most general proof of Eq.(\ref{1.2}).
For the special case $ \eps_n = \eps = \eps_0 $, we may also write
\beq
g_P ( x_1, \tau | x_0, 0 ) =
\langle x_1 | e^{ - i H \tau} P( n \eps) \cdots P (2 \eps ) P ( \eps ) | x_0 \rangle
\label{T0}
\eeq

Each of the above propagators may be represented by a path integral,
\bea
g_V ( x_1, \tau | x_0, 0 ) &=& \int {\cal D}x \exp \left( i \int_0^\tau dt \left[ \half m \dot x^2 + i V_0
\theta (-x) \right] \right)
\label{2.5}
\\
g_P ( x_1, \tau | x_0, 0 ) &=& \int_P {\cal D}x \exp \left( i \int_0^\tau dt \half m \dot x^2 \right)
\label{2.4}
\eea
where in both cases the paths are from $x(0)=x_0$ to $x(\tau) = x_1$ and in the second
case, Eq.(\ref{2.4}), $P$ denotes that the paths are restricted to be in the positive $x$-axis
at times $t = \eps_0+ (k-1) \epsilon$, $k=1,2 \cdots n$.

A closely related object that will be important is the restricted
propagator,
\beq
g_r ( x_1, \tau | x_0, 0 ) = \int_{x(t)>0} {\cal
D}x \exp \left( i \int_0^\tau dt \half m \dot x^2 \right)
\label{2.6}
\eeq
where again the paths are from $x(0)=x_0$ to
$x(\tau) = x_1$ but with $x(t)>0$ for all times in $[0, \tau]$.
This is clearly equivalent to $g_P$ in Eq.(\ref{2.4}), in the limit
$ n \rightarrow \infty $, $ \eps \rightarrow 0 $
with $ \tau $ constant. If we take the same
limit in the equivalent expression for $g_P$, Eq.(\ref{2.1}),
one obtains the following convenient operator form of the restricted
propagator:
\beq
g_r (\tau,0) = P \exp ( - i P H P \tau )
\label{2.7}
\eeq
The restricted propagator satisfies the Schr\"odinger equation in $x>0$
subject to the boundary conditions that it vanishes when either end
of the propagator sits on $x=0$.
For the free particle, considered here, one can easily solve for the
restricted propagator using the
method of images and the result is
\bea
g_r ( x_1, \tau | x_0, 0 ) &=& \left( \frac {m} { 2 \pi i \tau} \right)^{1/2}
\theta (x_1) \theta (x_0) \nonumber \\
& \times &
\left[
\exp \left(  \frac { i m (x_1 - x_0)^2 } { 2 \tau} \right)
- \exp \left(  \frac { i m (x_1 + x_0)^2 } { 2 \tau} \right) \right]
\label{2.8}
\eea
The restricted propagator, in any of the above forms, describes the regime
of ``Zeno dynamics'', in which all states are confined entirely to the Hilbert
subspace of states with support only in $x>0$ \cite{Sch2}.

The restricted propagator plays an important role here since not only does
$g_P \rightarrow g_r $
in the limit $ n \rightarrow \infty $, $ \eps \rightarrow 0 $
with $ \tau $ constant, but also $ g_V \rightarrow g_r $ as $V_0 \rightarrow \infty$.
It follows that $g_V$ and $g_P$ become arbitrarily close to each other for
sufficiently large $V_0$ and $n$, since they both tend to the same limit.
This is the underlying reason why we expect the approximate equivalence Eq.(\ref{1.2})
should hold.

The propagators $ g_P$ and $g_V$ may be
decomposed using the path decomposition expansion (PDX), in which
a propagator, regarded as a sum over paths,
is split into propagation corresponding to paths (or sections of paths)
entirely in $x>0$, paths entirely in $ x<0$ and paths starting
and ending on the boundary \cite{PDX,HaOr,Hal3}. There are two cases to consider.

We consider first the case in
which the initial and final points are on the same side of the surface in $x>0$.
The set of paths from initial to final point may be partitioned into paths that
never cross the origin, represented by the restricted propagator $g_r$, and paths
that always cross. The paths that cross have a first crossing at $t_1$
and a last crossing at $t_2$.
The corresponding PDX for any propagator $g$ has the form
\bea
g(x_1, \tau | x_0,0 ) &=&
\frac {1  } {4m^2} \int_{0}^{\tau} dt_2
\int_0^{t_2} dt_1
\ \frac {\partial g_r} {\partial x} (x_1, \tau | x, t_2) \big|_{x=0} \ g (0,t_2| 0,t_1)
\ \frac {\partial g_r } { \partial x} (x,t_1| x_0,0) \big|_{x=0}
\nonumber \\
&+&g_r ( x_1, \tau | x_0, 0 )
\label{2.9}
\eea
This is depicted in Figure 1. The first term in this expression, describing paths
which cross the origin, contains the derivative of $g_r$ because this corresponds
to paths which never cross $x=0$ but end or begin at $x=0$ (recalling that
$g_r$ itself vanishes if either end is on $x=0$) \cite{PDX}.

Eq.(\ref{2.9}) holds for both $g_P$ and
$g_V$, but the restricted part $g_r $ is the same in each case and equal to
Eq.(\ref{2.8}) above, since the restriction on paths defining $ g_P$ and
the presence of the complex potential in
$g_V$
in Eqs.(\ref{2.5}), (\ref{2.4}) are both redundant if $x(t)>0$.
Eq.(\ref{2.9}) may therefore be
simplified using
\beq
 \frac {\partial g_r } { \partial x} (x,t_1| x_0,0) \big|_{x=0}
 =  2 \frac {\partial g_f } { \partial x} (0,t_1| x_0,0)
\label{simp}
\eeq
which follows from Eq.(\ref{2.8}),
where $g_f$ denotes the free particle propagator.

It follows from the above that
$g_P$ and $g_V$ could differ only in terms of their propagation
along $x=0$, hence to prove the approximate equivalence of the propagators
Eqs.(\ref{2.2}), (\ref{2.1}), we need to prove that
\beq
g_P (0, t_2 | 0, t_1 ) \approx g_V (0, t_2| 0, t_1 )
\label{2.11}
\eeq
Note that, unlike $g_V (0,t_2|0,t_1)$, $ g_P (0, t_2 | 0, t_1 ) $ is
not covariant under time translation, that is,
\beq
g_P (0, t_2 | 0, t_1 ) \ne g_P (0, t_2 -t_1 | 0, 0 )
\label{cov}
\eeq
although it has an approximate covariance on timescales much greater than $\eps$,
as we will see below.

The second case is that in which the initial
and final points are on opposite sides of the surface, $x_1 < 0 $ and $x_0 > 0 $.
In this case we have for $g_V$,
\bea
g_V(x_1, \tau | x_0,0 ) &=&
\frac {1  } {4m^2} \int_{0}^{\tau} dt_2 \ \ e^{ - V_0 ( \tau - t_2) }
\ \int_0^{t_2} dt_1
\ \frac {\partial g_r} {\partial x} (x_1, \tau | x, t_2) \big|_{x=0}
\nonumber \\
& \times & \ g_V (0,t_2| 0,t_1)
\ \frac {\partial g_r } { \partial x} (x,t_1| x_0,0) \big|_{x=0}
\label{2.13}
\eea
since the last section of the paths is in $x<0$ where the complex potential acts.
For $g_P$, if $x_1 < 0$, since the paths must be in $x>0$ at the given discrete
set of times,
the last crossing time $t_2$ cannot be less than the last time $ \tau - \eps_n $
at which the projectors act, hence
\bea
g_P(x_1, \tau | x_0,0 ) &=&
\frac {1  } {4m^2} \int_{\tau - \eps_n}^{\tau} dt_2
\int_0^{t_2} dt_1
\ \frac {\partial g_r} {\partial x} (x_1, \tau | x, t_2) \big|_{x=0}
\nonumber \\
& \times &
\ g_P (0,t_2| 0,t_1)
\ \frac {\partial g_r } { \partial x} (x,t_1| x_0,0) \big|_{x=0}
\label{2.14}
\eea
Again we will have to prove that Eq.(\ref{2.11}) holds, but these two expressions
Eqs.(\ref{2.13}), (\ref{2.14}) also differ in the form of the $t_2$ integral.
However, since $ V_0 \eps \approx 1 $, the form of the exponential in
Eq.(\ref{2.13}) effectively squeezes $t_2$ to lie approximately
within $\eps $ of $\tau$, so we have
\beq
\int_{0}^{\tau} dt_2 \ \ e^{ - V_0 ( \tau - t_2) } \approx
\int_{\tau - \eps_n}^{\tau} dt_2
\eeq
and Eqs.(\ref{2.13}) and (\ref{2.14}) are approximately the same.

We see that both cases reduce to proving Eq.(\ref{2.11}).
The propagator along the boundary for the complex potential is known \cite{Car},
and is given by
\bea
g_{V}(0,t|0,0)&=&\left(\frac{m}{2\pi i
t}\right)^{1/2}\frac{(1-e^{-V_{0}t })}{V_{0}t }
\nonumber \\
&:=& \left(\frac{m}{2\pi i t}\right)^{1/2} f_V (t)
\label{CP}.
\eea
The main purpose of the remainder of this paper is to calculate
the propagator with projection operators along the boundary,
\beq
g_P ( 0, \tau | 0, 0 ) = \langle 0 | e^{-iH \epsilon_{n}} Pe^{-i H \epsilon}
\cdots e^{-i H \epsilon}P e^{-i H \eps_0} | 0 \rangle
\label{GP}
\eeq
and show that
the approximation Eq.(\ref{2.11}) holds, under conditions to be determined.
(Here, $ |0 \rangle $ denotes a position eigenstate $ | x \rangle $ at $x=0$).

In what follows, our main result is to show that, to a good approximation,
\beq
g_P ( 0, t | 0, 0 ) \approx \left(\frac{m}{2\pi i t}\right)^{1/2} f_P (t)
\label{FP1}
\eeq
where $f_P (t)$ is a kind of saw-tooth function -- a piecewise linear function
with peaks immediately followed by troughs at $t_k = \eps_0 + (k-1) \eps $,
$k= 1,2,\cdots$.
Approaching $t_k$ from below, there is a peak of value
\beq
f_P (t_k) = \frac {1} {k+1}
\label{peaks}
\eeq
and approaching $t_k$ from above there is a trough of half that size.
That is
\beq
f_P (t) = \frac {(t - t_{k-1})} {(k+1) \eps} + \frac { (t_{k} - t) } {2 k \eps}
\ \ \ \ {\rm for } \ \ \ \ t_{k-1} \le t < t_{k}, \ \ k = 2,3 \cdots
\label{FP2}
\eeq
and
\beq
f_P(t) = 1 \ \ \ \ {\rm for} \ \ \ \ 0 \le t < \eps_0
\eeq
The functions $f_P (t)$ and $f_V (t) $  are shown in Figure 2.
We see that $f_P (t)$
oscillates with period $\eps$ about $ f_V (t)$, as long as
we choose $V_0$ so that $f_V (t) $ lies between the peaks and troughs of $f_P (t)$.
That is,
for large $k$, we require that
\beq
\frac {1} {2k} < \frac{1} { V_0 t} < \frac {1} {k}
\eeq
Since $ t \approx k \eps $, $f_V (t)$ will
lie approximately midway between the peaks and troughs of $f_P (t)$ if
\beq
V_0 \eps \approx \frac {4} {3}
\eeq
Recalling that the propagator is attached through the PDX Eq.(\ref{2.9})
to an initial state, the oscillations, and hence the
differences between $g_P$ and $g_V$, will be smoothed out as long
as $\eps $ is chosen to be smaller than the timescale of variation
of the initial state. With some qualifications (discussed further in Section 8),
this timescale
is the Zeno time, $ 1 / \Delta H$.
The desired approximation Eq.(\ref{2.11}) will therefore hold in a time-averaged
sense, and
we thus have significant agreement
with the extended version of the original argument of Echanobe et al
described in Section 2.

In the following sections, we evaluate Eq.(\ref{GP}) and confirm the
form Eq.(\ref{FP2}) of $f_P(t)$.
In Section 4 we evaluate Eq.(\ref{GP}) exactly for the cases of one, two and
three projections.
We also show why in general the troughs of $f_P(t)$ are exactly half the size of the
peaks by considering the limit $\eps_n \rightarrow 0 $ of Eq.(\ref{GP}). In Section 5,
we use a lattice method to derive the magnitude of the peaks of
$f_P (t)$ for large $k$.
These results are substantiated
in Section 6, where we use an S-matrix expansion to derive some exact
results for a certain time-averaged version of Eq.(\ref{GP}).
In Section 7,
we fill in some of the gaps in these
regimes and approximations by computing Eq.(\ref{GP}) using
numerical methods.

We will find in the numerical and analytic solutions
that the function interpolating between the troughs
and peaks of $f_P(t)$ is not in fact a linear function in general. It is
interesting however, that, because of the slow time variation of the initial
state in comparison to $\eps$,
the precise form of this function turns out to be unimportant, and
this is what makes the problem more tractable than one might expect.
All that is important is the location of the peaks and troughs
of the saw-tooth function and the period of their oscillation.
It is this separation of timescales that also restores an approximate
time translation covariance to $g_P$, even though it does not hold exactly
in general, Eq.(\ref{cov}).

\section{Exact Analytic Results}

In this section we carry out an exact evaluation of Eq.(\ref{GP}) for the case
of one, two and three projections. For simplicity, we take the initial time interval
$\eps_0$ to be $\eps$ and for the case of three projections we are able to carry
out the calculation only in when the final time $\eps_n = \eps$.
We thus compute the objects
\beq
g_P (0,t|0,0) =
\begin{cases}
\langle 0 | e^{ - i H t} | 0 \rangle, &\mbox{if}  \ 0\le t < \eps \\
\langle 0 | e^{ - i H (t-\eps)} P e^{ - i H \eps} | 0 \rangle, &\mbox{if}  \ \eps \le t < 2 \eps \\
\langle 0 | e^{ - i H (t-2 \eps)} P e^{ - i H \eps} P e^{ - i H \eps} | 0 \rangle, &\mbox{if}  \ 2 \eps \le t < 3 \eps \\
\langle 0 | e^{ - i H \eps }P e^{ - i H \eps} P e^{ - i H \eps} P e^{ - i H \eps}
| 0 \rangle, &\mbox{if} \ t = 4 \eps
\end{cases}
\label{4E.1}
\eeq
We will show that
\beq
g_P (0,t|0,0) =
\begin{cases}
\left( \frac {m} {2 \pi i t} \right)^{\half} &\mbox{if}  \ 0\le t <  \eps \\
\half \left( \frac {m} {2 \pi i t} \right)^{\half} &\mbox{if}  \ \eps \le t < 2 \eps \\
\frac {1} {4} \left( 1 + \frac {2} {\pi} {\rm ArcTan} [(t-2 \eps) /t ]^{\half}  \right) \left( \frac {m} {2 \pi i t} \right)^{\half}
&\mbox{if}  \ 2 \eps \le t \ < 3 \eps \\
\frac {1} {4} \left( \frac {m} {2 \pi i t} \right)^{\half}
&\mbox{if}  \ t = 4 \eps
\end{cases}
\label{4E.2}
\eeq
thereby confirming the approximate form of the saw-tooth function, Eq.(\ref{FP2}), for the
first few projections. These expressions all have the property that the value
of $g_P(0,t|0,0)$ drops to half its value immediately after a projection and we show
why this is true in general at the end of this section.

To compute Eq.(\ref{4E.1}) it turns out to be
useful to consider more general objects, in which the projections may
be $P$, $ \bar P$ or the identity. We therefore consider the object
\bea
g_{P,C}(0,t |0,0) &=&\int_{C}dx_{1}...dx_{n}\left(\frac{m}{2\pi i \e}\right)^{n/2} \left(\frac{m}{2\pi i (t - n \eps)}\right)^{1/2}
\nonumber \\
& \times & \exp\left(\frac{imx_{1}^{2}}{2\e}+\frac{im}{2\e}\sum_{k=1}^{n-1}(x_{k}-x_{k+1})^{2}+\frac{imx_{n}^{2}}{2(t - n \eps)}\right)
\eea
Here $C=\{+,-,0...\}$ symbolically stands for the integration ranges of the $x_{k}$, eg $C=\{+,-,0...\}$
means $0<x_{1}<\infty, -\infty<x_{2}<0, -\infty<x_{3}<\infty$ etc.
Performing a Wick rotation, $\e\to-i\e$ and changing variables yields
\bea
g_{P,C}(0,-i t|0,0)&=&\left(\frac{m}{2\pi }\right)^{1/2}\int_{C}\frac { dy_{1}...dy_{n} } { (\pi \eps)^{n/2} (t- n \eps)^{1/2} }
\nonumber \\
& \times &
\exp\left(-\frac {y_{1}^{2}} {\eps} -\sum_{k=1}^{n-1}\frac {(y_{k}-y_{k+1})^{2}} {\eps} -\frac {y_{n}^{2}} { (t - n \eps)}
\right)\\
\nonumber \\
& := & \left(\frac{m}{2\pi }\right)^{1/2} T_C ( \eps, t)
\label{tc}
\eea

The cases of no projection and one projection are trivially evaluated, and we easily
obtain the first two equations in Eq.(\ref{4E.2}).
For the case of two projections, we need to evaluate the object
\beq
T_{++}(\e_{1},\e_{2},\e_{3})=\int_{0}^{\infty}\frac{dy_{1}dy_{2}}{\pi \sqrt{\e_{1} \e_{2} \e_{3}}}
\exp\left(-\frac{y_{1}^{2}}{\e_{1}}-\frac{(y_{1}-y_{2})^{2}}{\e_{2}}-\frac{y_{2}^{2}}{\e_{3}}\right)
\eeq
(a slight generalization of the $T_C$ defined in Eq.(\ref{tc})).
Changing variables to $r=y_{1}$, $y=y_{2}/y_{1}$ gives,
\bea
T_{++}(\e_{1},\e_{2},\e_{3})&=&\int_{0}^{\infty}\frac{dr dy}{\pi\sqrt{\e_{1}\e_{2}\e_{3}}}r
\exp\left(-r^{2}\left\{\frac{1}{\e_{1}}+\frac{(1-y)^{2}}{\e_{2}}+\frac{y^{2}}{\e_{3}}\right\}\right)\nonumber\\
&=&\int_{0}^{\infty}\frac{dy}{2\pi}\frac{1}{(\e_{2}\e_{3}+\e_{1}\e_{3}(1-y)^{2}+\e_{1}\e_{2}y^{2})}\nonumber\\
&=&\int_{0}^{\infty}\frac{dy}{2\pi}\frac{1}{a+by+cy^{2}}
\eea
Where $a=\e_{3}(\e_{1}+\e_{2})$, $b=-2\e_{1}\e_{3}$, $c=\e_{1}(\e_{2}+\e_{3})$.
Now use the change of variables,
\beq
u=\frac{b}{2\sqrt{c}}+\sqrt{c}y
\eeq
to obtain
\beq
T_{++}(\e_{1},\e_{2},\e_{3})=\frac{1}{2\pi}\sqrt{\frac{\e_{2}}{\e_{1}\e_{3}c}}
\int_{\frac{b}{2\sqrt{c}}}^{\infty}du\frac{1}{u^{2}+\frac{4ac-b^{2}}{4c}}
\eeq
Noting that $4ac-b^{2}>0$, and using the standard integral \cite{GrRy}
\beq
\int du \frac{1}{u^2+\a^{2}}=\frac{1}{\a}{\rm ArcTan}\left(\frac{u}{\a}\right)
\eeq
we find finally
\bea
T_{++}(\e_{1},\e_{2},\e_{3})
&=&\frac{1}{4\pi\sqrt{(\e_{1}+\e_{2}+\e_{3})}}\left(\pi+2{\rm ArcTan}
\left(\sqrt{\frac{\e_{1}\e_{3}}{\e_{2}(\e_{1}+\e_{2}+\e_{3})}}\right)\right)
\label{ex1.7}
\eea
Setting $\e_1 = \e_2 = \eps $ and $\e_3 = t - 2 \e $ it follows that
\beq
g_P (0, t |0, 0) =
\frac {1} {4} \left( 1 + \frac {2} {\pi} {\rm ArcTan} [(t-2 \eps) /t ]^{\half}  \right)
\left( \frac {m} {2 \pi i t} \right)^{\half}
\eeq
for $ 2 \eps \le t < 3 \eps $,
so we confirm the third equation in Eq.(\ref{4E.2}).

It will be useful for the three projection case below to record the following related results.
A similar analysis to that above yields
\bea
T_{+-}(\e_{1},\e_{2},\e_{3})&=&\int_{0}^{\infty}dy_{1}\int_{-\infty}^{0}dy_{2}
\frac{1}{\pi \sqrt{\e_{1} \e_{2} \e_{3}}}\exp\left(-\frac{y_{1}^{2}}{\e_{1}}-
\frac{(y_{1}-y_{2})^{2}}{\e_{2}}-\frac{y_{2}^{2}}{\e_{3}}\right)\nonumber\\
&&=\frac{1}{4\pi\sqrt{(\e_{1}+\e_{2}+\e_{3})}}\left(\pi-2{\rm ArcTan}
\left(\sqrt{\frac{\e_{1}\e_{3}}{\e_{2}(\e_{1}+\e_{2}+\e_{3})}}\right)\right)
\label{ex1.8}
\eea
In particular we then have
\bea
T_{++}&=& T_{++}(\e,\e,\e)=\frac{1}{3\sqrt{3 \e}}
\\
T_{+-}&=&\frac{1}{6\sqrt{3\e}}
\\
T_{+0}&=&\frac{1}{2\sqrt{3\e}}= \half T_{00}
\eea

We are able to evaluate the three projection case only in the situation where all time intervals
are equal, so we set $ t - n \eps = \eps $ in the definition of $T_C$ in  Eq.(\ref{tc}).
In this case, $T_{C}$ possesses a number of helpful symmetries. The first is ``reflection'' symmetry:
if we define $-C$ by the string obtained by letting $(+\to -), (-\to+) ,(0\to 0)$ then
\beq
T_{-C}=T_{C}\label{prop1}
\eeq
The second symmetry is ``time reversal'': if we define $\tilde C$ by the string
obtained by reversing the order of $C$, then we have
\beq
T_{\tilde C}=T_{C}\label{prop2}
\eeq
In addition we have the simple property that
\beq
T_{...0...}=T_{...+...}+T_{...-...}\label{prop3}
\eeq
These properties imply the following for the three projection case.
On the one hand we have,
\beq
T_{+++}=T_{++0}-T_{++-}=T_{++0}-T_{0+-}+T_{-+-}
\eeq
but we also have,
\beq
T_{+++}=T_{+0+}-T_{+-+}=T_{+0+}-T_{-+-}
\eeq
Combining these expressions gives,
\beq
2T_{+++}=T_{+0+}+T_{++0}-T_{0+-}
\eeq
For each of the objects on the right, we may carry out the full range integral,
to leave ourselves with an object of the form $T_{C}(\e_{1},\e_{2},\e_{3})$ computed above.
For example we have that $T_{+0+}=T_{++}(\e,2\e,\e)$.
Each of these objects may then be evaluated using Eq.(\ref{ex1.7}) and Eq.(\ref{ex1.8}),
and combined to give,
\beq
T_{+++}=\frac{1}{4\sqrt{4\e}}
\eeq
It follows that
\beq
g_P (0, t |0,0 ) = \frac {1} {4} \left( \frac {m} {2 \pi i t} \right)^{\half}
\eeq
when $t= 4 \eps $, thus confirming the fourth equation in Eq.(\ref{4E.2}).

To end this section, we confirm our claim in Section 3 that the function $g_P (0,t|0,0)$
drops to half its value immediately after a projection.
On the face of it, this involves
interpreting the expression $ \langle 0 | P $ which is ambiguous, so must be defined
by a limiting procedure (where, recall $ \langle 0 |$ denotes $ \langle x |$
at $x=0$).
We thus consider the limit
$ \eps_n \rightarrow 0 $ in the expression
\bea
g_P ( 0, \tau | 0, 0 ) &=& \langle 0 | e^{-iH \epsilon_{n}} Pe^{-i H \epsilon} P
\cdots e^{-i H \epsilon}P e^{-i H \eps_0} | 0 \rangle
\nonumber \\
&=& \int dy \ \langle 0 | e^{-iH \epsilon_{n}} P e^{-i H \epsilon} | y \rangle \langle y |  P
\cdots e^{-i H \epsilon}P e^{-i H \eps_0} | 0 \rangle
\label{GP4}
\eea
We seek to show that the result obtained by this limit
is half the result obtained when the final projection is absent.
We write
\beq
P = \half ( P + \bar P)  + \half ( P - \bar P)
\eeq
so that
\beq
\langle 0 | e^{-iH \epsilon_{n}} P e^{-i H \epsilon} | y \rangle
= \half \langle 0 | e^{-iH \epsilon_{n}} e^{-i H \epsilon} | y \rangle +
\half \langle 0 | e^{-iH \epsilon_{n}} (P - \bar P)  e^{-i H \epsilon} | y \rangle
\eeq
The first term on the right-hand side yields the sought after factor of a half as $\eps_n \rightarrow 0 $,
so it remains to show that the second term is zero. We have
\bea
\langle 0 | e^{-iH \epsilon_{n}} (P - \bar P)  e^{-i H \epsilon} | y \rangle
&=& \int_0^{\infty} dx \ \langle 0 | e^{-iH \epsilon_{n}} | x \rangle
\left( \langle x | e^{ - i H \eps } | y \rangle -  \langle - x | e^{ - i H \eps } | y \rangle \right)
\label{4E.23}
\eea
where we have used the fact that
\beq
\langle 0 | e^{-iH \epsilon_{n}} | x \rangle = \langle 0 | e^{-iH \epsilon_{n}} | -x \rangle
\eeq
(for the free particle, considered here).
As $\eps_n \rightarrow 0 $, $ \langle 0 | e^{-iH \epsilon_{n}} | x \rangle \rightarrow \delta (x)$
and the pair of terms in brackets on the right-hand side cancel in Eq.(\ref{4E.23}),
so we obtain zero as required.

\section{Asymptotic Limit}

We now consider the asymptotic evaluation of $g_P (0, \tau | 0, 0 )$ for large $n$
in the case $\eps_n = \eps = \eps_0$. This will give the values of the peaks
of $f_P(t)$, Eq.(\ref{peaks}).

Recall that in the limit $ n \rightarrow \infty $ and $ \eps \rightarrow 0 $
with $ \tau = (n+1) \eps $ fixed, $g_P (x_1, \tau | x_0, 0 )$
tends to the restricted propagator $g_r (x_1, \tau | x_0, 0) $.
To obtain the asymptotic form of $g_P (0, \tau | 0, 0 )$ for large $n$ we therefore
need to determine the lowest non-trivial correction to this result. It is clear
from the definition Eq.(\ref{2.1}) of $g_P$ that close to the limit,
we have the general form
\beq
g_P (x_1, \tau | x_0, 0 ) =  g_r (x_1, \tau | x_0, 0 ) + \eps g_1 (x_1, \tau | x_0, 0 )
+ O (\eps^2)
\label{5.1}
\eeq
for some function $g_1 (x_1, \tau | x_0, 0)$.
Since the restricted propagator vanishes if either $x_1 = 0 $ or $ x_0 = 0 $, the object
we need to calculate is
\beq
g_1 (0,\tau|0,0) = \lim_{\eps \rightarrow 0, n \rightarrow \infty} \frac { g_P (0, \tau | 0, 0 ) } {\eps}
\label{limit}
\eeq

A standard and convenient way to do this is to rotate to imaginary time $\tilde \tau $
(we use tilde to denote Euclideanized time)
and then write $g_P (x_1, \tilde \tau | x_0, 0 )$  as a
Euclidean path integral. This is
then defined in terms of the continuum limit of probabilities of random walks on a space time lattice
of temporal spacing $ \Delta \tilde \tau $ and spatial spacing $\eta $.
The details of this construction are very conveniently
given by Hartle \cite{Har}
so we will give only the briefest account here.
Using this language,
$g_P (0, \tilde \tau | 0, 0 )$ is then the continuum limit of the object
$ (2 \eta)^{-1} u_P  ( 0, \tilde \tau | 0, 0) $, where $ u_P  ( 0, \tilde \tau | 0, 0)  $ is
the probability for a random walk on the lattice starting at the origin and
ending at
the origin at time $ \tilde \tau$, with the restriction that the walker is
in the positive $x$-axis at the intermediate times $ \tilde \eps, 2 \tilde \eps, 3 \tilde \eps
\cdots$ (where $ \tilde \eps \ge \Delta \tilde \tau $).

The calculation of $u_P$ defined in this way, for values of $\tilde \eps $ generally
greater than the lattice spacing $\Delta \tilde \tau$,
is in fact a known problem in combinatorics called the tennis ball problem. It
appears to have  a formal solution, but this solution is too implicit for us to
extract a useful result \cite{tennis}.

Fortunately, however, for the purposes of calculating
the limit Eq.(\ref{limit}) the results of Ref.\cite{Har}
are sufficient.
For this case, we set
$\tilde \eps = \Delta \tilde \tau$
and $ u_P( 0, \tilde \tau | 0, 0) $ is then
the probability for a random walk from the origin to itself, with the restriction
that the walker is in the positive $x$-axis at {\it every} intermediate step.
On a finite lattice Hartle's calculations give the result,
\beq
\frac {1} {2 \eta} u_P( 0, \tilde \tau | 0, 0) =
\left( \frac{ m } { 2 \pi \tilde \tau} \right)^{1/2} \ \frac { \tilde \eps} {\tilde \tau}
\eeq
to leading order for small $\tilde \eps$, $\eta$ \cite{Har}. We may use this result
to compute the limit Eq.(\ref{limit}), which, continued back to real time, is
\beq
g_1 (0, \tau | 0, 0 ) = \left(\frac{m}{2\pi i
\tau}\right)^{1/2} \frac { 1} {\tau}
\eeq
Through Eq.(\ref{5.1}), this confirms Eq.(\ref{peaks})
for large $k$.

\section{A Time Averaged Result}

We have argued that at the peak values of $ g_P (0,\tau|0,0) $ with $n$
projections, we have the approximate result
\beq
\langle 0 | e^{-iH \tau } P (n \eps) \cdots P (2 \eps ) P ( \eps )
| 0 \rangle
\approx \left(\frac{m}{2\pi i \tau}\right)^{1/2} \frac {1} {n+1}
\label{TA}
\eeq
We showed above that this is exact for $n=1,2,3$ and true asymptotically
for large $n$.
Some interesting exact
results for any $n$ may be obtained by considering a more general version of this object
in which the projections are not restricted to act at the given set of
times, $ \eps, 2\eps, 3 \eps \cdots$.

On the one hand, Eq.(\ref{CP}) may be expanded as a power series in powers of
$V_0$:
\beq
g_{V}(0,\tau|0,0)=\left(\frac{m}{2\pi i
\tau}\right)^{1/2}\sum_{n=0}^{\infty} \frac{ (-1)^n }{ (n+1)! } V_0^n \tau^n
\label{T1}
\eeq
On the other hand, the evolution operator with complex potential may be expanded
in the usual S-matrix expansion,
\bea
e^{-i(H-iV)\tau} &=& e^{-iH\t} \ T \exp \left( -\int_{0}^{\t}dt V(t)\right)
\nonumber \\
&=& e^{-iH \tau}\sum_{n=0}^{\infty}\frac{(-1)^{n}}{n!}\int_{0}^{\tau}dt_{n}...\int_{0}^{\tau}dt_{1}
T [V(t_{n})...V(t_{1})]
\label{T2}
\eea
where $T$ denotes time ordering,  $V (t) = V_0 \bar P (t) $ and $\bar P(t) = e^{ i H t } \bar P e^{ - i H t} $.
It follows that
\bea
g_{V}(0,\tau|0,0) &=& \langle 0 | \exp \left(-i(H-iV)\tau \right) | 0 \rangle
\nonumber \\
&=& \sum_{n=0}^{\infty}\frac{(-1)^{n}}{n!} V_0^n \int_{0}^{\tau}dt_{n}...\int_{0}^{\tau}dt_{1}
\ \langle 0 | e^{-iH\t} T [\bar P(t_{n})...\bar P(t_{1})] | 0 \rangle
\label{T3}
\eea
Equating powers of $V_0$ in Eqs.(\ref{T1}) and (\ref{T3}) and writing out the time ordering
explicitly, we deduce that
\beq
\frac{n!} {\tau^n}  \int_{0}^{\tau}dt_{n} \int_0^{t_n} dt_{n-1}...\int_{0}^{t_2}dt_{1}
\ \langle 0 | e^{-iH\t}  \bar P(t_{n})... \bar P(t_{1}) | 0 \rangle
= \left(\frac{m}{2\pi i
\tau}\right)^{1/2} \frac {1} {n+1}
\label{T4}
\eeq
We would like to write this in a form involving $P$, instead of $\bar P$. We introduce
the reflection operator
\beq
R = \int dx \ | x \rangle \langle -x |
\eeq
and note that $ \bar P = R P R $, the Hamiltonian $H$ commutes with $R$, and $ R | 0 \rangle = | 0 \rangle $
(where, recall, $ | 0 \rangle $ denotes $ | x \rangle $ at $x=0$). It follows that
\beq
\langle 0 | e^{-iH\t}  \bar P(t_{n})... \bar P(t_{1}) | 0 \rangle
= \langle 0 | e^{-iH\t}  P(t_{n})... P(t_{1}) | 0 \rangle
\eeq
so that Eq.(\ref{T4}) holds with all the $\bar P$'s replaced with $P$'s.
Noting that
\beq
\frac{n!} {\tau^n}  \int_{0}^{\tau}dt_{n} \int_0^{t_n} dt_{n-1}...\int_{0}^{t_2}dt_{1} = 1
\eeq
we see that Eq.(\ref{T4}) is of the desired general form, Eq.(\ref{TA}), but time-averaged
over the times of the projections.

The question is now to what extent the time-averaged expression on the left-hand
side
of Eq.(\ref{T4}) is close to Eq.(\ref{TA}). We may take this further in two different
ways.

First, note that for a real-valued function $f$ of $n$ variables, we have
the mean value theorem
\beq
\frac{n!} {\tau^n}  \int_{0}^{\tau}dt_{n} \int_0^{t_n} dt_{n-1}...\int_{0}^{t_2}dt_{1}
f(t_n, t_{n-1} \cdots t_1) =  f( \xi_n, \xi_{n-1} \cdots \xi_1 )
\label{T5}
\eeq
for some set of numbers $\xi_n \ge \xi_{n-1} \ge \cdots \ge \xi_1 $ in the interval
$[0,\tau]$. The integral in Eq.(\ref{T4}) is easily made into a real integral
over a real-valued function by analytic continuation, and
we therefore deduce the exact result
\beq
\langle 0 | e^{-iH\t}  P(\xi_{n})...P(\xi_{1}) | 0 \rangle
= \left(\frac{m}{2\pi i
\tau}\right)^{1/2} \frac {1} {n+1}
\eeq
for {\it some} set of projection times $\xi_n \ge \xi_{n-1} \ge \cdots \ge \xi_1 $
in the interval $[0, \tau]$.

Second, we may expand the integrand in the left-hand side of Eq.(\ref{T4}) about
the values $ k \eps $, to get some insight into why these particular values
have any special significance. We have
\bea
\langle 0 | e^{-iH\t}  P(t_{n})...P(t_{1}) | 0 \rangle
&=& \langle 0 | e^{-iH\t}  P(n \eps)...P(\eps ) | 0 \rangle
\nonumber \\
&+&
\sum_{k=1}^n \ (t_k - k \eps ) \  \frac {\partial} { \partial t_k} \langle 0 | e^{-iH\t}
P(t_{n})...P(t_{1}) | 0 \rangle \big|_{t_k = k \eps}
\nonumber \\
&+& \cdots
\label{5.8}
\eea
Inserting in Eq.(\ref{T4}), and noting that
\beq
\frac{n!} {\tau^n}  \int_{0}^{\tau}dt_{n} \int_0^{t_n} dt_{n-1}...\int_{0}^{t_2}dt_{1} \ t_k = k \eps
\eeq
(where, recall, $ \tau = (n+1) \eps $)
we see the first order term in the expansion Eq.(\ref{5.8}) averages to zero in
Eq.(\ref{T4}). This is clearly only
true for expansion about the special values $t_k = k \eps $. We therefore obtain
the desired result Eq.(\ref{TA}) up to second order corrections. This suggests that
the values $t_k = k \eps $ are significant because they
give the best approximation to the average in Eq.(\ref{T4}).
These results give evidence that Eq.(\ref{peaks}) holds approximately
for all $n$, including the intermediate values not covered in the previous
two sections.


\section{Numerical Results}

To support the analytic results described in the previous sections
we evaluate Eq.(\ref{GP}) numerically for up to $20$ projections and confirm
the conjectured form Eq.(\ref{FP1}), (\ref{FP2}).
It is convenient to define a sequence of functions
$F_n (t,x)$ defined by
\beq
F_0 (t,x) = \langle x | e^{ - i H t } | 0 \rangle
\eeq
for $ 0 \le t < \eps $ and
\beq
F_n (t,x) = \langle x | e^{ - i H ( t - n \eps ) } \left( P e^{ - i H \eps} \right)^n | 0 \rangle
\eeq
for $ n \eps \le t \le (n+1) \eps $ where $n=1,2,3 \cdots $. The desired object Eq.(\ref{GP})
(with, for convenience, $\eps_0 = \eps$), is then given by
\beq
g_P (0,t|0,0) = F_n (t,0)
\eeq
The sequence $F_n (t,x)$ may be calculated using the recursion relation,
\beq
F_n (t,x) = \int_0^\infty dy \ \langle x | e^{ - i H (t - n \eps ) } | y \rangle \ F_{n-1} (n \eps, y)
\eeq
Using a new time coordinate $s$ defined by $ t = s \eps $, rotating to imaginary time,
and defining $\tilde F_n ( s, x) = F_n ( - i t, x ) $,
we have
\beq
\tilde F_n (s, x ) = \left( \frac{m} { 2 \pi ( s - n ) } \right)^{\half}
\int_0^\infty dy \ \exp \left( - \frac { m(x-y)^2 } { 2(s-n) } \right)\ \tilde F_{n-1} (n, y)
\label{7.5}
\eeq
for $ n \le s \le n+1 $.

We have attempted to find an approximate analytic solution to Eq.(\ref{7.5}) for large $n$,
but without success. However, a numerical solution is straightforward and yields
all the information we require.
This was done using a simple mid-point rule. The lattice
size was chosen to be of order $10^{-3}$, and the results were checked
for robustness against changes in lattice size.

The numerical result for $f_P (t) $ (defined in Eq.(\ref{FP1}) in terms of $g_P (0,t|0,0)$)
is plotted in Figure 3, along with our claimed approximate analytic expression
for $f_P (t)$, Eq.(\ref{FP2}). We see that there is excellent agreement.
The values at the peaks and troughs appear to agree perfectly.
The only small
discrepancy is that the interpolating functions between the peaks and troughs
are not exactly linear. This discrepancy is only noticeable for intermediate
values of $n$ and in any event since, as argued, the curve is effectively
averaged over time in the PDX, this discrepancy is insignificant. We therefore
find substantial numerical confirmation for our our main result, Eqs.(\ref{FP1}), (\ref{FP2}).

The apparently perfect agreement of numerical results with the approximate analytic expression
Eq.(\ref{FP2}) at the peaks and troughs is striking. We wonder if the approximate
analytic expression is in fact exact at these points, but we have not been able
to prove this, except for the cases $n=1,2,3$ and for large $n$.

A useful way of seeing even more precisely the relationship between $f_P
(t)$ and $ f_V (t)$ is to define the function
\beq
S(t) = \frac { f_P (t) } { f_V (t) } - 1
\label{7.6}
\eeq
so that
\beq f_P (t) = ( 1 + S(t) ) f_V (t)
\eeq
This is plotted in Figure 4.
It is a simple function oscillating around
zero between $ \pm 1/3 $ with period $\eps$. In terms of $S(t)$, the relationship between $g_P$ and $g_V$ then has the particularly
simple form
\beq
g_P (0,t|0,0) = ( 1 + S(t) ) g_V (0,t|0,0)
\label{7.7}
\eeq
This relationship is perhaps the most concise summary of the sought-after connection between the propagators.

\section{Timescales}

We now give a more detailed explanation as to the timescales
involved in proving the approximate equivalence of $g_V$ and $g_P$.
We broadly expect that the appropriate timescale is the
Zeno time of the initial state, $ t_Z = 1 / \Delta H$. However, we have derived a very precise
connection between evolution in the presence of a complex potential
and evolution with projection operators so we are in a position
to investigate the specific way in which an initial state may discriminate
between these two types of evolution.

We have argued that the equivalence boils down to proving
the equivalence of propagation along the boundary, Eq.(\ref{2.11}),
and we have shown that $ g_P (0,t|0,0) $ oscillates around $ g_V (0,t|0,0)$
with period $\eps$. Suppose we have an initial state $\psi (x,0)$.
Let us consider the change in the wave function
arising from replacing $ g_V $ with $g_P$
using
the PDX, Eq.(\ref{2.9}). It
is
\beq
\delta \psi ( x_1, \tau) = \frac {1  } {m^2} \int_{0}^{\tau} dt_2
\int_0^{t_2} dt_1
\ \frac {\partial g_f} {\partial x} (x_1, \tau | x, t_2) \big|_{x=0} \ \delta g (0,t_2| 0,t_1)
\ \frac {\partial \psi} { \partial x } (0,t_1)
\label{8.1}
\eeq
where
\beq
\delta  g (0,t_2| 0,t_1) =  g_V (0,t_2| 0,t_1) -  g_P (0,t_2| 0,t_1)
\eeq
and we also used Eq.(\ref{simp}).
The important part of this expression is the $t_1$ integral, which we
expect will be small if the initial state is sufficiently slowly varying
in time.

The results
of Section 3 and 8 show that $ \delta g (0,t|0,0)$ oscillates around zero
with period $\eps$. In particular, Eq.(\ref{7.7}) shows that
\beq
\delta g (0,t|0,0) = S(t) g_V (0,t|0,0)
\eeq
The explicit form of $S(t)$ is given in Figure 4, but its important
qualitative feature is its oscillation with period $\eps$, so
for simplicity, $S(t)$ may
be loosely modelled by the function $ \sin (2 \pi t/\eps)$.
To be definite
we take the initial state $\psi$ to be a Gaussian wave packet, so we have
\beq
\psi (x,t) = N \exp \left( - \frac { ( x - q - pt/m)^2 } {4 \sigma^2} + i p x - i E t \right)
\eeq
where $E = p^2 / 2m$, $N$ is a normalization factor, and we have ignored wave packet spreading
effects in evolving the state. For such a state the Zeno time is $t_Z = m \sigma / p $.
In Eq.(\ref{8.1}) the differentiation of $ \psi $ produces a prefactor which does not contribute
to the leading order evaluation of the time integral, and if the range of integration is much
greater than $\eps$, we obtain the order of magnitude result
\beq
\int dt_1 \ \delta g (0,t_2| 0,t_1)
\ \frac {\partial \psi} { \partial x } (0,t_1) \sim \exp \left( - \frac {t_Z^2} {\eps^2} (E \eps - 1)^2 \right)
\label{8.4}
\eeq
If $ E \eps \ll 1 $, the right-hand side is clearly very small if $ \eps \ll t_Z $.
If $ E \eps > 2 $, the right-hand side if bounded from above by $ \exp ( -  {t_Z^2}/ {\eps^2} ) $
so again will be small if $ \eps \ll t_Z $. Hence the Zeno time of the initial state
is indeed the timescale controlling the validity of the approximation, as expected.
Note also that Eq.(\ref{8.4}) goes to zero as $\eps \rightarrow 0$, as it must, since
$g_P$ and $g_V$ become exactly equal (and equal to $g_r$) in this limit.

The only problematic case is that in which the initial state has energy $ E \sim 1 / \eps$. In this
case, the right-hand side of  Eq.(\ref{8.4}) is not necessarily small and the approximation
may fail. This is not surprising since it is the case in which the oscillations in time of the
incoming state are comparable to the temporal spacing of the projections, so that the state
can ``see'' the difference between the complex potential and projections at a discrete
set of times.

Hence, apart from the above exception, Eq.(\ref{2.11}) holds for $ \eps \ll t_Z$.
As outlined in the Introduction, reflection in the complex potential
is negligible if $V_0 \ll E $ which is equivalent to  $ \eps \gg 1/E $.
Therefore there is an interesting regime, namely
\beq
\frac {1}  {E} \ll \eps \ll \frac {1} {\Delta H}
\eeq
in which the approximate
equivalence Eq.(\ref{2.11}) holds, yet there is negligible reflection. This
regime is important in, for example, study of the arrival time problem using
complex potentials \cite{HaYe1}.
This issue will considered in more detail elsewhere \cite{Yea}.
(See also Ref.\cite{Sch} for an interesting
discussion of timescales in the Zeno effect.)

\section{Summary and Discussion}

This paper was physically motivated by the desire
to understand the effect of periodically acting projections
onto the positive $x$-axis for a free particle, Eq.(\ref{1.1}).
A valuable way to proceed is to use the conjectured
relationship Eq.(\ref{1.2}) with a complex potential first put forward by Echanobe et al.
This connection, together with known results on scattering,
establishes the timescale under which significant reflection occurs
in Eq.(\ref{1.1}).
We noted that the arguments for the relationship Eq.(\ref{1.2}) are only heuristic and
there is scope for a more substantial proof.

We proved Eq.(\ref{1.2}) by considering the associated propagators Eqs.(\ref{2.2}), (\ref{2.1}).
We noted that an approximate equivalence between these propagators
is expected since both propagators tend to the restricted
propagator $g_r (x_1, \tau | x_0,0)$
in the limits
$ \eps \rightarrow 0$, $ n \rightarrow \infty $ and $ V_0 \rightarrow \infty$.
The path decomposition expansion reduced the proof of equivalence
of these propagators to the simpler case of propagation between points
lying on the origin, Eq.(\ref{2.11}). The propagator along
the origin for the complex potential $g_V (0,t|0,0)$ is already known, Eq.(\ref{CP}), so the bulk of the proof
was to derive the analogous result for the propagator with
projections, $g_P( 0 , \tau | 0, 0 )$. Our main result was to prove
that this propagator
has the approximate form Eqs.(\ref{FP1}), (\ref{FP2}), which we proved
using a variety of analytic and numerical methods. In effect, the main achievement
of this paper has therefore been to obtain a good approximate analytic expression
for the propagator $g_P$ that appears in Eq.(\ref{1.1}).

We found that $g_P(0,t|0,0)$ oscillates with period $\eps$
around $g_V (0,t|0,0)$ as
long as $V_0 \eps \approx 4/3 $, a result most concisely summarized in Eq.(\ref{7.7}).
The approximate equivalence Eq.(\ref{2.11})
of these propagators then holds in a time-averaged sense as long as the
timescale $\eps$ between projections is much smaller than the Zeno
time of the initial state, $ 1 / \Delta H $ (but may fail in the special
case when the incoming state is peaked about energy $ E \sim 1/ \eps$).
These conditions agree in essence with those of an extended version
of the results of Echanobe et al, with an advantage over their results
that a definite relationship between $V_0$ and $\eps$ is obtained.
We noted that their restriction Eq.(\ref{1.3}) relating
$ V_0$ and $\eps$ is in fact stronger than required and the equality
$ V_0 \eps \approx 4/3 $ derived here gives the best approximate
equivalence between $g_P $ and $g_V$.

In addition to helping establish the timescales for reflection in Eq.(\ref{1.1}),
the connection Eq.(\ref{1.2}) has been of use in investigating the near-Zeno
regime of Eq.(\ref{1.1}) \cite{Ech,Muga2,Hal4}. In the limit $\eps \rightarrow 0$,
the Zeno limit, motion in Eq.(\ref{1.1}) is confined entirely to states with
support only in $x>0$ and is described by the restricted propagator Eq.(\ref{2.7}).
It is of interest to explore the nature of the dynamics very close to this
regime, i.e., for small but finite $\eps$. This can be achieved by examining
the form of the propagator with complex potential $g_V$ for large $V_0$.
In this regime, one can calculate, for example, the (unnormalized) probability density
for crossing the origin between $ \tau $ and $ \tau + d \tau$,
which is found to be
\bea
\Pi (\tau) &=& - \frac { d N } { d \tau }
\nonumber \\
&=& \frac { 2} { m^{3/2} V_0^{1/2}}
\langle \psi_f (\tau) |\hat p \delta ( \hat x ) \hat p
| \psi_f (\tau) \rangle
\label{6}
\eea
Here, $N (\tau)$ is the survival probability (given by the norm of the state
evolved in the presence of the complex potential)
and $ | \psi_f (\tau) \rangle $ is the freely evolved wave function.
This is proportional to the average kinetic energy density at the origin.
As anticipated it goes to zero as $V_0 \rightarrow \infty$ since there is
total reflection. However, it is useful to define a normalized distribution,
\bea
\Pi_N (\tau) &=& \frac { \Pi ( \tau) } {  \int_0^\infty d t \Pi (t )  }
\nonumber \\
&=& \frac { 1 } { m \langle p \rangle }
\langle \psi_f (\tau) |\hat p \delta ( \hat x ) \hat p
| \psi_f (\tau) \rangle
\label{7}
\eea
where
$\langle p \rangle $ is the average momentum in the initial wave packet \cite{Ech,Muga2,Hal4}.
Interestingly, this is now independent of the complex potential despite this being
the regime of strongly-acting measurement. This regime will be explore further in future
publications \cite{Yea}.

Finally, we comment on the possible generality of the connection Eq.(\ref{1.2}).
We first note that it may in fact be written
\beq
e^{ - i H \eps} Pe^{-i H \epsilon} \cdots P e^{-i H \eps}
\approx \exp \left( - i H \tau - V_0 \bar P \tau \right)
\label{9.1}
\eeq
where $\bar P = 1 - P$. We have proved Eq.(\ref{9.1}) for the case
in which the projections are onto the positive $x$-axis, but it seems
reasonably clear that the relationship will hold for projections onto
any region $[a,b]$ of the $x$-axis (as long as it is not too small) and indeed
for regions of configuration space in a many-dimensional model.
Such a potential has been used recently in the decoherent histories analysis
of quantum cosmological models \cite{Hal7}).
Moreover, although the proof of Eq.(\ref{9.1}) given in this paper
relied heavily on the fact that $P$ projects onto position,
the form Eq.(\ref{9.1}) and the heuristic
argument for it in Section 2 do not rely on the particular form of $P$.
We therefore conjecture that Eq.(\ref{9.1}) may hold for a wider
variety of projection operators, not just projectors onto position.
This will be pursued elsewhere.

\acknowledgements

We are grateful to Carl Bender for useful discussions and to Adolfo del Campo
for helpful comments on the first draft of this paper.



\bibliography{apssamp}

\vfil\eject

\epsfxsize=10cm
\epsfbox{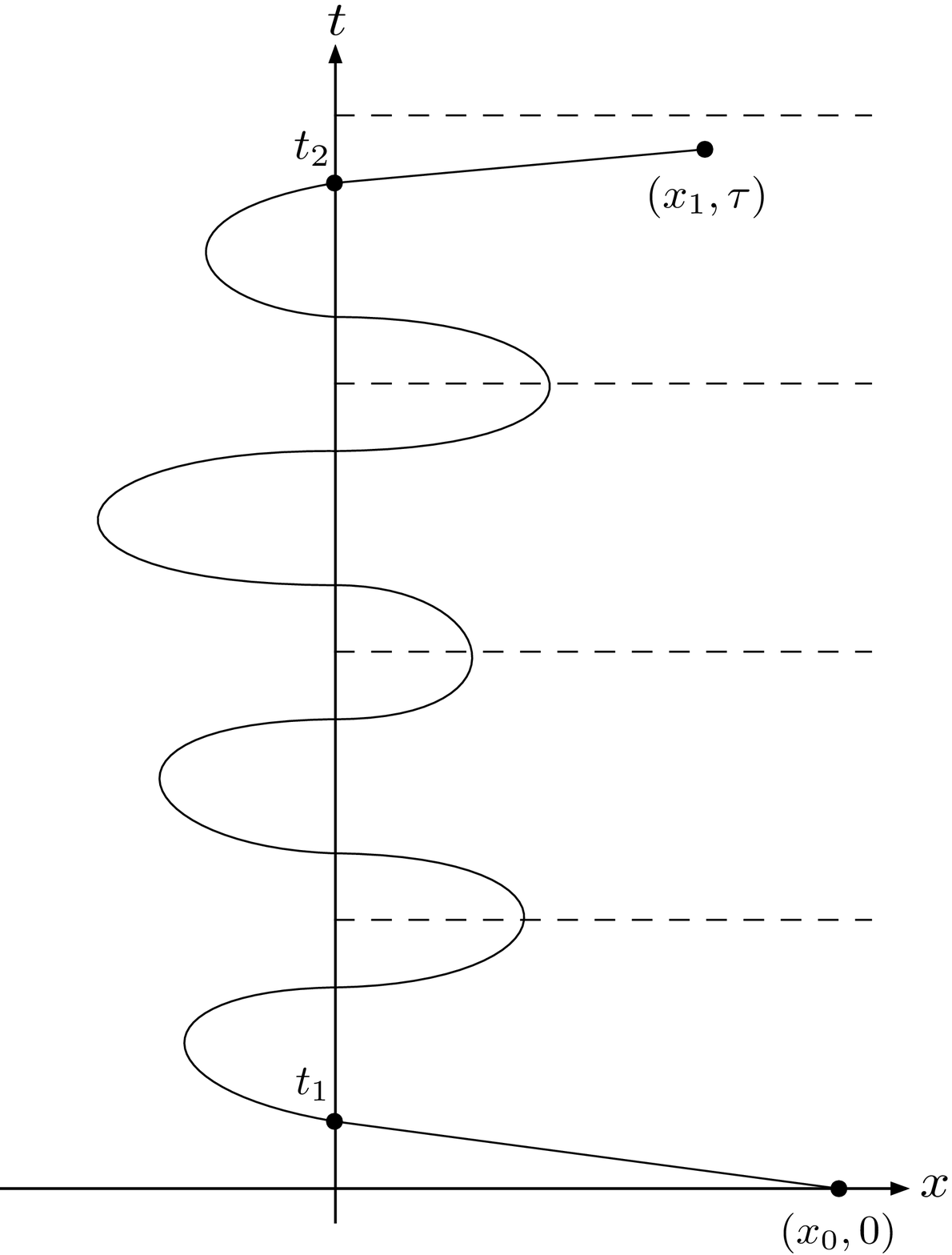}

\noindent{\bf Figure 1.} The path decomposition expansion for $g_P$.
Any path from $ x_0 > 0 $
at $t=0$ to a final point $x_1 > 0 $  at $t=\tau$ which crosses $x=0$
has a first crossing of $x=0$ at $t_1$ and
a last crossing at $t_2$.
The crossing part of the propagator from $(x_0,0)$ to $(x_1, \tau)$ may therefore
be decomposed
into three parts: (A) restricted propagation entirely in $x>0$, (B) propagation
starting and ending on $x=0$ with the restriction that the paths
are in $x>0$ at a discrete set of times, and (C) restricted propagation entirely in $x>0$.
The corresponding path decomposition expansion formula is given in Eq.(\ref{2.9}).
(The $g_r$ term corresponds to paths which never cross $x=0$).

\vfil\eject

\epsfxsize=15cm
\epsfbox{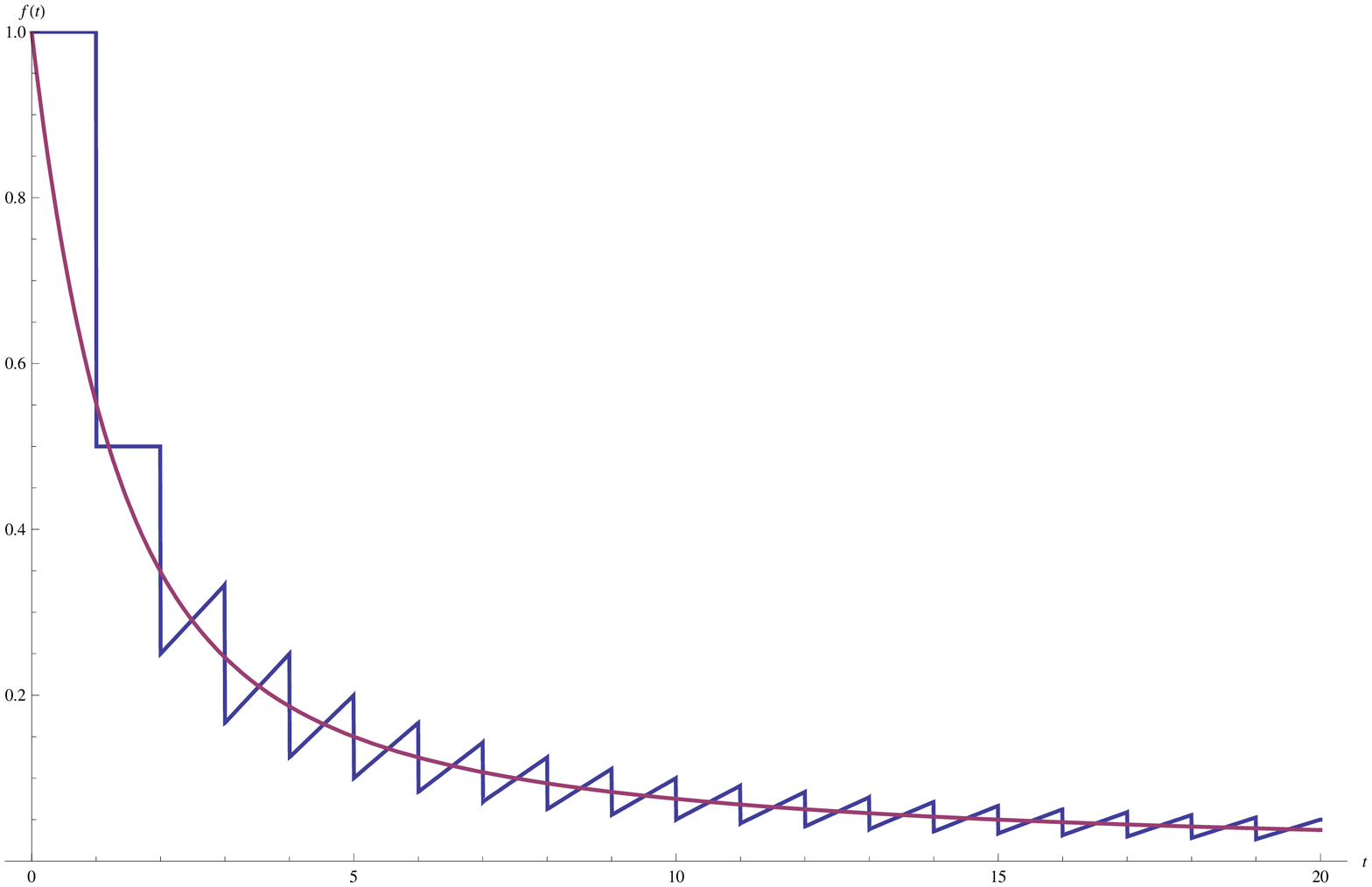}

\noindent{\bf Figure 2.} A plot of the functions $f_V(t)$ and $ f_P (t)$,
defined for the complex potential by Eq.(\ref{CP})
with $ V_0 \eps = 4/3$,
and for intermittent projectors by Eqs.(\ref{FP1}), (\ref{FP2}). (The time scaling
is such that $\eps = 1$ in the plot, so that the peaks occur at integer values of $t$).
We see that $f_P (t)$
oscillates around $f_V (t)$ with period $\eps$ so the two functions are
equal in a time-averaged sense, when integrated against functions which
are slowly varying on a timescale $\eps$.

\vfil\eject

\epsfxsize=15cm
\epsfbox{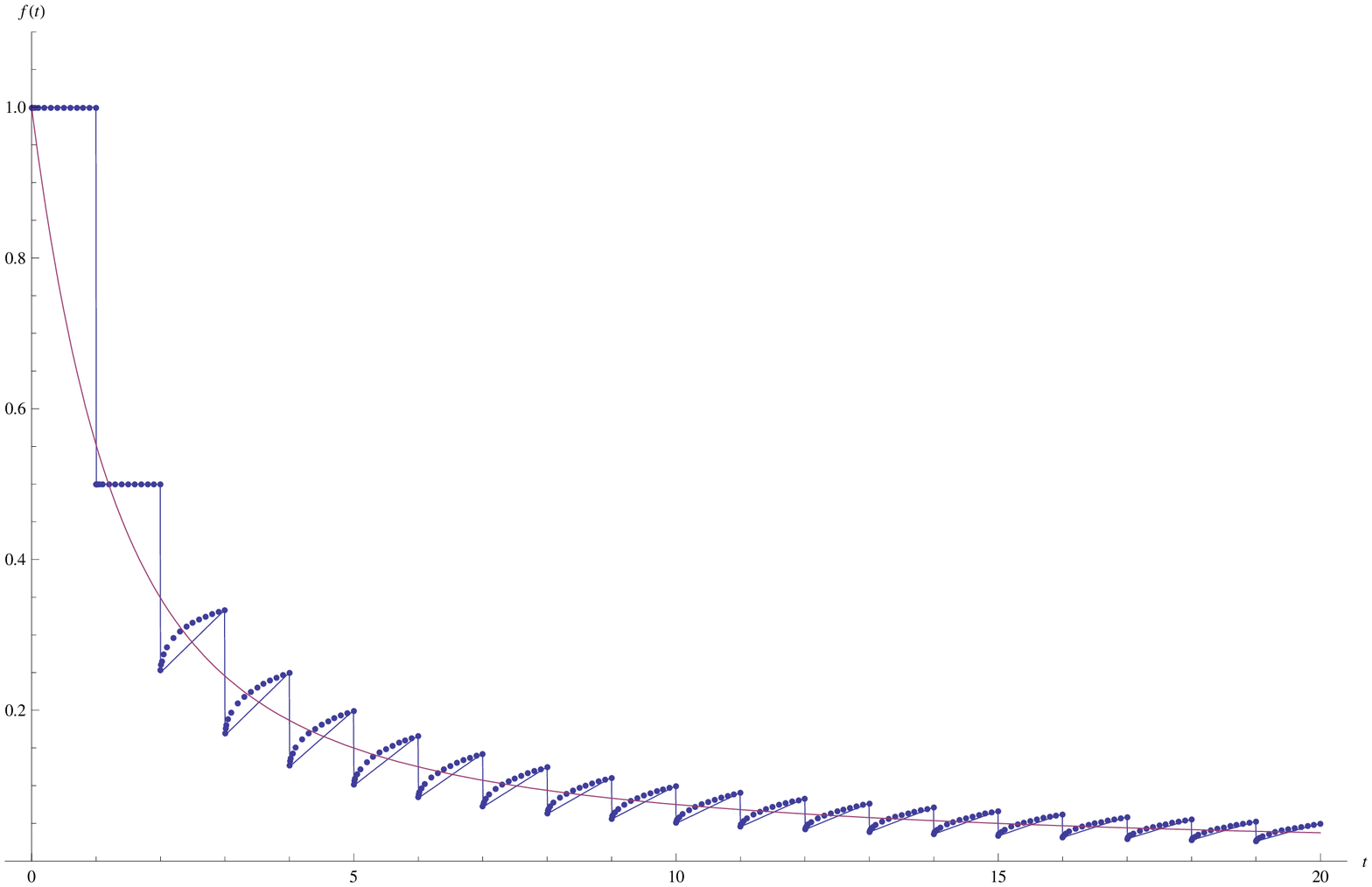}

\noindent{\bf Figure 3.} A plot of the numerical calculation of $f_P (t)$
(the set of dotted lines),  and the conjectured analytic expression for
$f_P(t)$, Eqs.(\ref{FP1}), (\ref{FP2}) (the set of straight lines).
(Again we use a time scaling such that $\eps = 1$).
There is excellent agreement between analytic and numerical results at the
peaks and troughs of the functions. The numerical result
shows some departure from exact linear behaviour between the peaks and troughs
but this is insignificant as argued in the main text.
A plot of $f_V (t)$, the smooth curve, is also given for reference.

\vfil\eject

\epsfxsize=15cm
\epsfbox{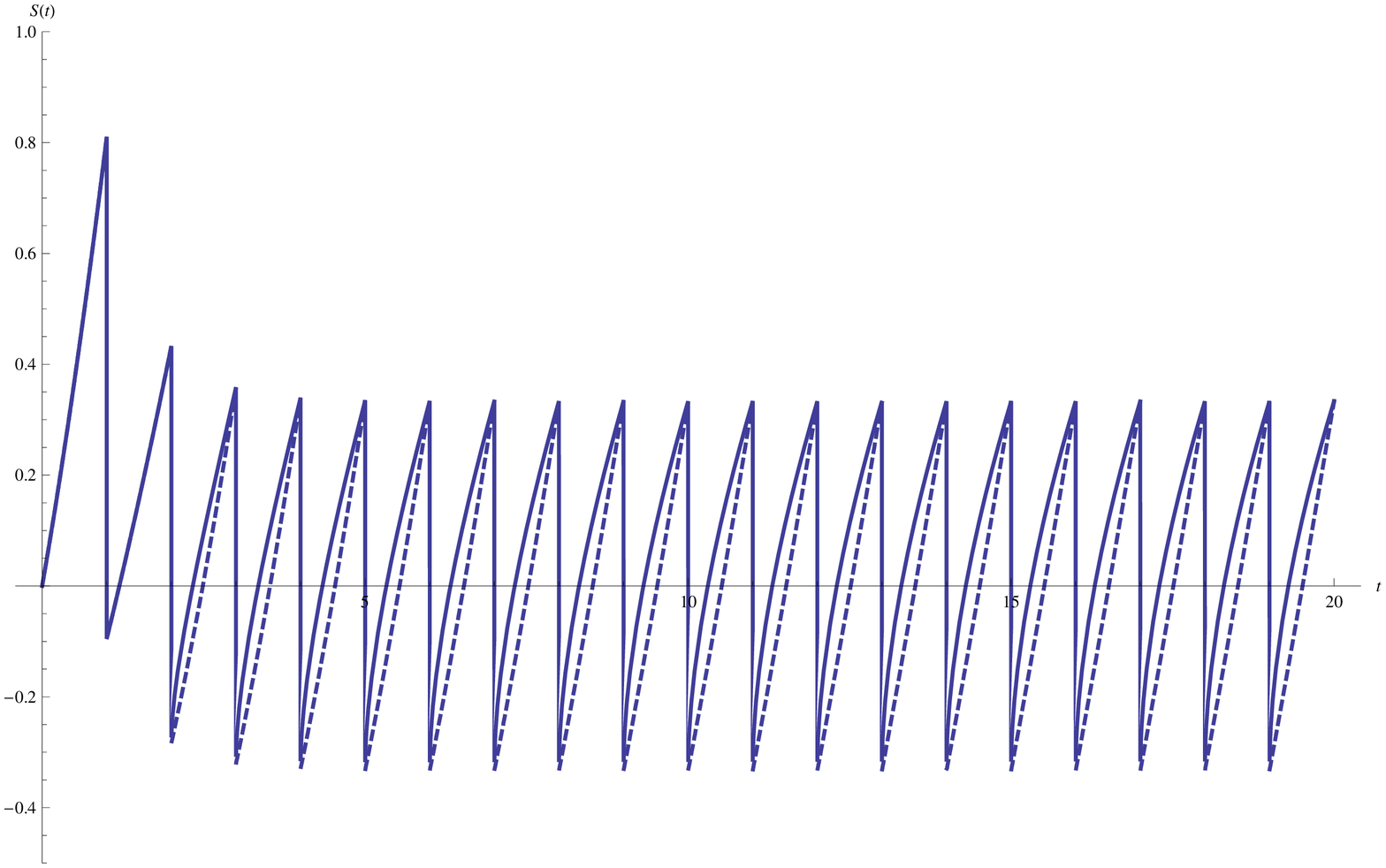}

\noindent{\bf Figure 4.}  A plot of the function $S(t)$ defined in Eq.(\ref{7.6}),
describing the oscillations of $g_P (0,t|0,0)$ around $g_V (0,t|0,0)$. The bold line
represents the numerical calculation and the dashed line the analytic result
(from Eqs.(\ref{CP}), (\ref{FP2})). After the first few oscillations, it
oscillates between $ \pm 1/3$ (approximately).

\end{document}